\documentclass[aps,twocolumn,floatfix]{revtex4}

\usepackage{times}
\usepackage{graphics}% Include figure files
\usepackage{epsfig,calc}% Include figure files
\usepackage{bm}  %boldmath
\usepackage{amsmath,amssymb,chemarr}

\begin{document}

\title{Phase behavior and rheology of sticky rod-like particles}

\author{F. Huang$^1$, R. Rotstein$^1$, K. E. Kasza$^2$, N. T. Flynn$^3$, and S. Fraden$^1$}

\affiliation{
$^1$Department of Physics, Brandeis University, Waltham, MA 02454, USA\\
$^2$Department of Physics \& DEAS, Harvard University, Cambridge, MA
02138 USA\\
$^3$Department of Chemistry, Wellesley College, Wellesley, MA
02481, USA }

\date{\today}

\begin{abstract}
We construct colloidal ``sticky'' rods from the semi-flexible filamentous fd virus and temperature-sensitive polymers poly(N-isopropylacrylamide) (PNIPAM). The phase diagram of fd-PNIPAM system becomes independent of ionic strength at high salt concentration and low temperature, i.e. the rods are sterically stabilized by the polymer. However, the network of sticky rods undergoes a sol-gel transition as the temperature is raised. The viscoelastic moduli of fd and fd-PNIPAM suspensions are compared as a function of temperature, and the effect of ionic strength on the gelling behavior of fd-PNIPAM solution is measured. For all fluidlike and solidlike samples, the frequency-dependant linear viscoelastic moduli can be scaled onto universal master curves.

\end{abstract}

\maketitle

\section{Introduction}
The phase behavior of a fluid of rod-like particles interacting through short range repulsion has been well described at the second virial coefficient level by Onsager~\cite{Onsager49} who demonstrated that this system exhibits an isotropic-nematic ($I$-$N$) phase transition. Examples of colloidal liquid crystals range from minerals~\cite{Davidson05} to viruses~\cite{Dogic06} and in many examples experiments and theory agree. In this paper we modify the fd particle in order to introduce interparticle attractions. One approach to introduce attractions has been through "depletion attraction"~\cite{Asakura58} in which rods and polymers are mixed resulting in an attractive potential of mean force. Several theoretical works have incorporated depletion attraction into the Onsager theory~\cite{Warren94,Lekkerkerker94} and a simulation has also been performed~\cite{Bolhuis97}. These studies predict a widening of the biphasic $I$-$N$ gap. These results are in qualitative agreement with the measured $I$-$N$ transition in mixtures of boehmite rods and polystyrene polymers and mixtures of charged semiflexible fd virus and dextran polymers~\cite{Buitenhuis95,Bruggen00,Dogic04}.

For the case of direct interparticle attraction, theory also predicts that the width of the $I - N$ coexistence widens abruptly with increasing attraction~\cite{Flory56}. However, in experiments with the semiflexible polymer, PBG, experiments show that a gel phase supersedes the $I-N$~\cite{Miller74}.

In this work, we consider the effect of direct attractions on the phase behavior of colloidal rod-like particles. As a model colloidal rod we use aqueous suspensions of filamentous semiflexible bacteriophage fd. Suspensions of fd have been previously shown to exhibit an $I$-$N$ transition in agreement with theoretical predictions for semiflexible rods interacting with a salt dependent effective hard rod diameter $D_\text{eff}$~\cite{Tang95}. Although fd forms a cholesteric phase, the difference in free energy between the cholesteric and nematic phases is much smaller than that between the isotropic and nematic phases. Hence we refer to the cholesteric phase as the nematic phase in this paper.

%Since the phase behavior of fd virus depends weakly on temperature~cite{?},
We have developed a temperature sensitive aqueous suspension of colloidal rods. Specifically,  thermosensitive poly(N-isopropylacrylamide) polymers (PNIPAM) are covalently linked to the virus major coat protein pVIII. Solutions of poly(NIPAAM) exhibit a lower critical solution temperature (LCST) behavior in water. Below its LCST of $32^{\circ}$C, poly(NIPAAM) is readily soluble in water, while above its LCST the polymer sheds much of its bound water and becomes hydrophobic, which leads to collapse of the coil, attraction between polymers, and phase separation~\cite{Gennes79,Schild92}.

While most of the experiments to date exploring the role of attraction have been performed on systems where the attraction is due to the polymer induced depletion~\cite{Buitenhuis95,Bruggen00,Dogic04}, our experimental system has its advantages in that the  strength of attractive interaction can be finely tuned by adjusting the temperature of the solution. The range of the attraction can be controlled by the size of grafted polymers.

We explore the behavior of suspensions of fd-PNIPAM particle as a function of temperature. A sol-gel transition is found for both the isotropic and nematic phase and is studied in detail with dynamic light scattering (DLS) and rheometry. As the system can be driven reversibly from a fluidic state to a gel state, it provides a versatile model system to study the fundamental properties of entangled and crosslinked networks of semiflexible polymers.

%The excluded volume and solvent effects act oppositely and it is far from clear which feature, attraction or excluded volume will dominate

%The viscoelasticity of networks of semiflexible polymers has been a subject of intense experimental and theoretical investigation. Examples of semiflexible polymers include synthetic helical polypeptides such as poly(benzyl glutamate) (PBLG), aromatic polyamides such as Kevlar, and various biopolymers such as F-actin, microtubule and filamentous phage. F-actin is an essential component in the cytoskeletal networks that provide the mechanical stability to cells, and play a critical role in cell motion. A myriad of actin regulatory proteins can bind and crosslink actin filaments to create rigid actin network, which shows interesting and complex viscoelastic behaviors ref?. Recent efforts have been made to elucidate the role of these actin binding proteins ref?.

\section{Materials and Methods}
\subsection{Preparation of fd-PNIPAM complexes}
Bacteriophage fd is a rodlike semiflexible polymer of length $L=880$ nm, diameter $D=6.6$ nm, molecular weight $1.64\times10^7$ dalton, surface charge density 7$e^-$/nm at $p$H = 8.2 and persistence length between 1 and 2 $\mu$m~\cite{Fraden95,Khalil07}. There are approximately 2700 major coat proteins helically wrapped around the phage genome of a single-stranded DNA. fd virus is grown and purified as described elsewhere~\cite{Maniatis89}. The virus concentration is determined by UV absorption at 269 nm using an extinction coefficient of 3.84 cm$^{2}$/mg on a spectrophotometer (Cary-50, Varian, Palo Alto, CA).

About 30 mg NHS terminated PNIPAM (Polymer Source Inc., Quebec, Canada) is mixed with 800 $\mu$l of 24 mg/ml fd solution for 1 h in 20 mM phosphate buffer at $p$H = 8.0. The reaction product is centrifuged repeatedly to remove the excess polymers. The PNIPAM bound fd virus is stored in 5 mM phosphate buffer at $4^{\circ}$ for future use.
Using a differential refractometer (Brookhaven Instruments, Holtsville, NY) at $\lambda=620$nm, the refractive index increment, $(dn/dc)$, has been measured to estimate the degree of coverage of fd-PNIPAM complex~\cite{Grelet03}, and there are $336\pm60$ PNIPAM polymers grafted on each virus.

\subsection{Dynamic light scattering}
In a homodyne light scattering experiment, the time correlation function of the scattered light intensity is acquired,
\begin{eqnarray}
G_I(q,t)=\frac{\langle I(q,0)I(q,t)\rangle}{\langle I(q)\rangle^2}
\end{eqnarray}
This can be related to the correlation function of the electric field by the Siegert relation ~\cite{Berne76},
\begin{eqnarray}
G_E(q,t)=\sqrt{G_I(q,t)^2-1}
\end{eqnarray}
where
\begin{eqnarray}
G_E(q,t)=\frac{\langle E^*(q,0)E(q,t)\rangle}{\langle I(q,t)\rangle}
\end{eqnarray}
An effective diffusion coefficient can be defined by the first cumulant
\begin{eqnarray}
D_{\text{eff}}(q)=\Gamma(q)/q^2
\label{eq:Deff}
\end{eqnarray}
where
\begin{eqnarray}
\Gamma=-\frac{d}{dt}[\ln G_E(q,t)]_{t\rightarrow 0}
\end{eqnarray}
Here the $D_{\text{eff}}(q)$ reflects the different types of motion associated with the rod-like fd-PNIPAM particle, including translation, rotation and bending motion.

A light scattering apparatus (ALV, Langen, Germany) consisting of a computer controlled goniometer table with focusing and detector optics, a power stabilized 22 mW HeNe laser ($\lambda=633$ nm), and an avalanche photodiode detector connected to an $8\times8$ bit multiple tau digital correlator with 288 channels was used to measure the correlation function. The temperature of the sample cell in the goniometer system is controlled to within $\pm$0.1$^{\circ}$C.

To remove dusts and air bubbles in the fd-PNIPAM solution, the sample is passed through a 0.45 $\mu$m filter and centrifuged at 3000 rpm for 15 min before each measurement. The correlation function of the scattered light intensity was measured by the correlator at a scattering angle of $90^{\circ}$. The particle concentration ranges from $2c^*$ to $4c^*$ with the critical concentration $c^*=1$ particle/$L^3$ or 0.04 mg/ml.

\subsection{Rheological characterization of fd-PNIPAM suspensions}
The rheological measurements were carried out on a stress-controlled rheometer (TA Instruments, New Castle, DE) using a stainless steel cone/plate tool ($2^{\circ}$ cone angle, 20 mm cone diameter). The gap is set at 70 $\mu$m. The torque range is 3 nN$\cdot$m to 200 mN$\cdot$m, and the torque resolution is 0.1 nN$\cdot$m. The temperature control is achieved by using a Peltier plate, with a range of $-20^{\circ}$C to $200^{\circ}$C and an accuracy of $\pm$0.1$^{\circ}$C.

The storage and loss moduli, $G'(\omega)$ and $G''(\omega)$ respectively, are measured as a function of frequency by applying a small amplitude oscillatory stress at a strain amplitude $\gamma=0.03$. A strain sweep is conducted prior to the frequency sweep to ensure the operation is within the linear viscoelastic regime.

\begin{figure}
\epsfig{file=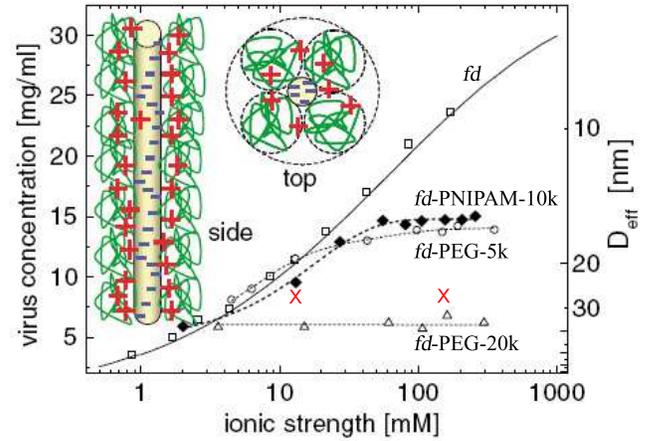,width=8.6cm} \caption{
    (Color online) Isotropic-nematic coexistence concentration of PNIPAM-coated fd virus as a function of ionic strength (solid symbol). The lines indicate the highest concentration for which the isotropic phase is stable. Plotted together as a comparison is data on wide-type fd, fd-PEG-5k and fd-PEG-20k taken from~\cite{Dogic01}. As the salt concentration increases, the fd-PNIPAM system transitions from an electrostatically-stabilized suspension to a sterically-stabilized suspension. This is schematically demonstrated by the cartoon of fd-PNIPAM particle with $D^\text{electrostatic}_\text{eff}<D^\text{polymer}_\text{eff}$. We measure the rheological properties of samples (cross symbol) in these two regimes.
}
\label{fd-PNIPAM_coex}
\end{figure}

\section{Results and Discussion}
Onsager~\cite{Onsager49} first predicted that there is an I-N phase transition in suspensions of hard rods when the number density of rods $c$ reaches $\frac{1}{4}c\pi L^2D=4$. Since the fd virus is charged, it's necessary to account for the electrostatic interaction by substituting the bare diameter $D$ with an effective diameter $D_{\text{eff}}$ which is larger that $D$ by an amount roughly proportional to the Debye screening length. As the solution ionic strength increases, $D_{\text{eff}}$ decreases and eventually approaches $D$. Fig.~\ref{fd-PNIPAM_coex} reveals the I-N coexisting concentrations as a function of ionic strength for bare fd~\cite{Tang95}, fd-PEG~\cite{Dogic01} and fd-PNIPAM particles respectively. All measurements are made at room temperature at which water is a good solvent for PNIPAM polymer. Similar to fd-PEG the I-N phase boundary of fd-PNIPAM is independent of ionic strength at high ionic strength. The physical picture is that $D_{\text{eff}}$ decreases with increasing ionic strength. Once $D_{\text{eff}} < D + D_{\text{poly}}$ the interparticle interactions are dominated by steric repulsion of the grafted PNIPAM and not electrostatic repulsion. For fd-PNIPAM the transition from electrostatic to polymer stabilized interactions occurs at $D_{\text{eff}} \sim 17$nm corresponding to a polymer diameter $D_{\text{poly}} = 10$nm.
%From the plateau value of coexistence concentration, we can calculate the effective diameter of fd-PNIPAM particle~ref{Dogic01} $D_{eff}=???$, and hence it's possible to estimate the effective size of attached PNIPAM molecule: $D_{bare}+4R_g=???$.

We study the phase behavior of fd-PNIPAM in response to temperature changes. We prepare samples in isotropic (9.6 mg/ml) and nematic (21 mg/ml) phase at $I=55$~mM. At room temperature, both isotropic and nematic samples are transparent viscous fluids. The nematic sample exhibits birefringence under cross polarizers while the isotropic sample does not. As the temperature is increased to $T=40^{\circ}$C, the samples rapidly turn into viscoelastic gels. These behaviors can be observed by simply tilting the vial, and observing the formation of a weight-bearing gel. As the temperature returns to room temperature, the samples flow like fluids again. The entire process can be repeated multiple times, which indicates a reversible sol-gel transition. This observation can be interpreted as the result of increased attraction among PNIPAM monomers leading to the collapse of PNIPAM coils into globules at elevated temperature and thus leading to an  attraction between the fd-PNIPAN rods.

We load the above mentioned samples into glass capillaries which are subsequently sealed with flame. The samples are placed in a heat block at $40^{\circ}$C, and monitored with polarizing microscopy for up to a week. No phase separation has been observed for both the isotropic and nematic samples, which remain in their respective phases. This is qualitatively different from the theoretical results that the addition of attraction could lead to phase separation~\cite{Lekkerkerker94}. We speculate the sticky rods at high temperature could be kinetically arrested in a non-equilibrium state and therefore do not phase separate during the course of the experiment.

\begin{figure}
\epsfig{file=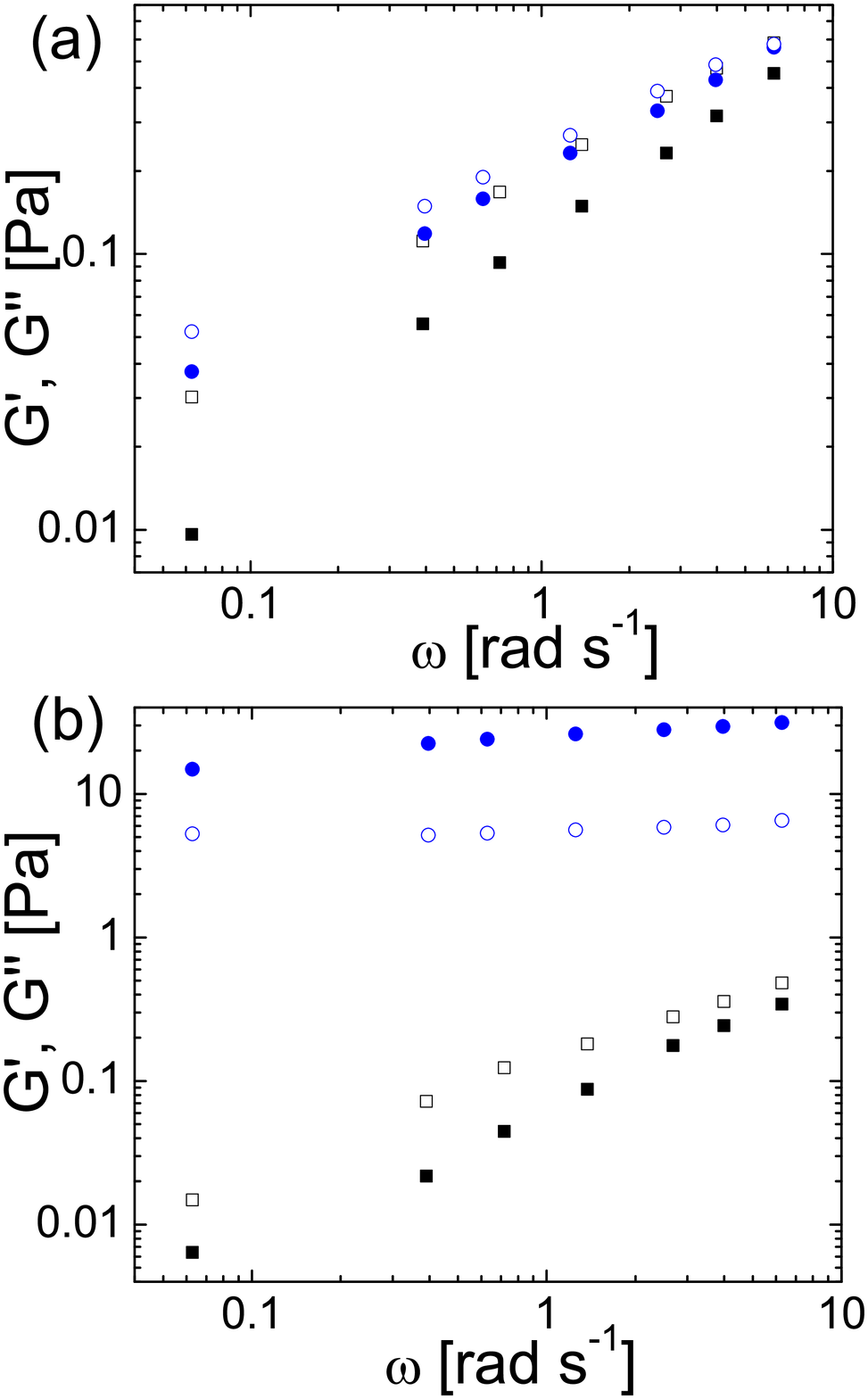,width=8.6cm} \caption{
    Storage modulus (solid symbol) and loss modulus (open symbol) of fd and fd-PNIPAM suspensions at two different temperatures. (a) $T=24^{\circ}$C, (b) $T=38^{\circ}$C. The concentration of the sample is about 8~mg/ml. The solution ionic strength is 155 mM.
}
\label{fd&fd-PN}
\end{figure}

\begin{figure}
\epsfig{file=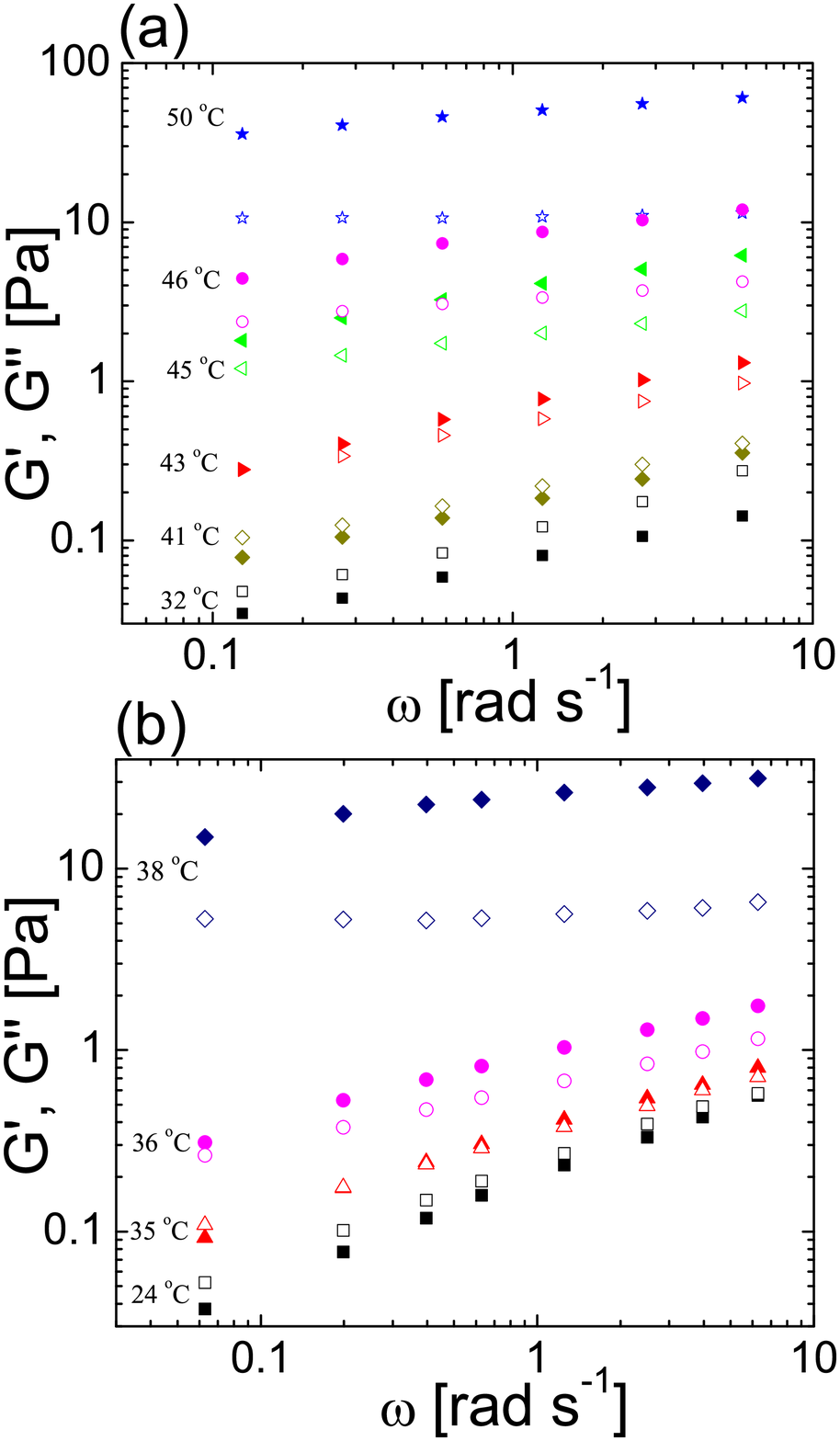,width=8.6cm} \caption{
    Storage modulus (solid symbol) and loss modulus (open symbol) of fd-PNIPAM solution as a function of temperature.  Ionic strength $I=$ (a) 13~mM, (b) 155~mM. The concentration of the sample is about 8~mg/ml.
}
\label{fSweep}
\end{figure}

\begin{figure}
\epsfig{file=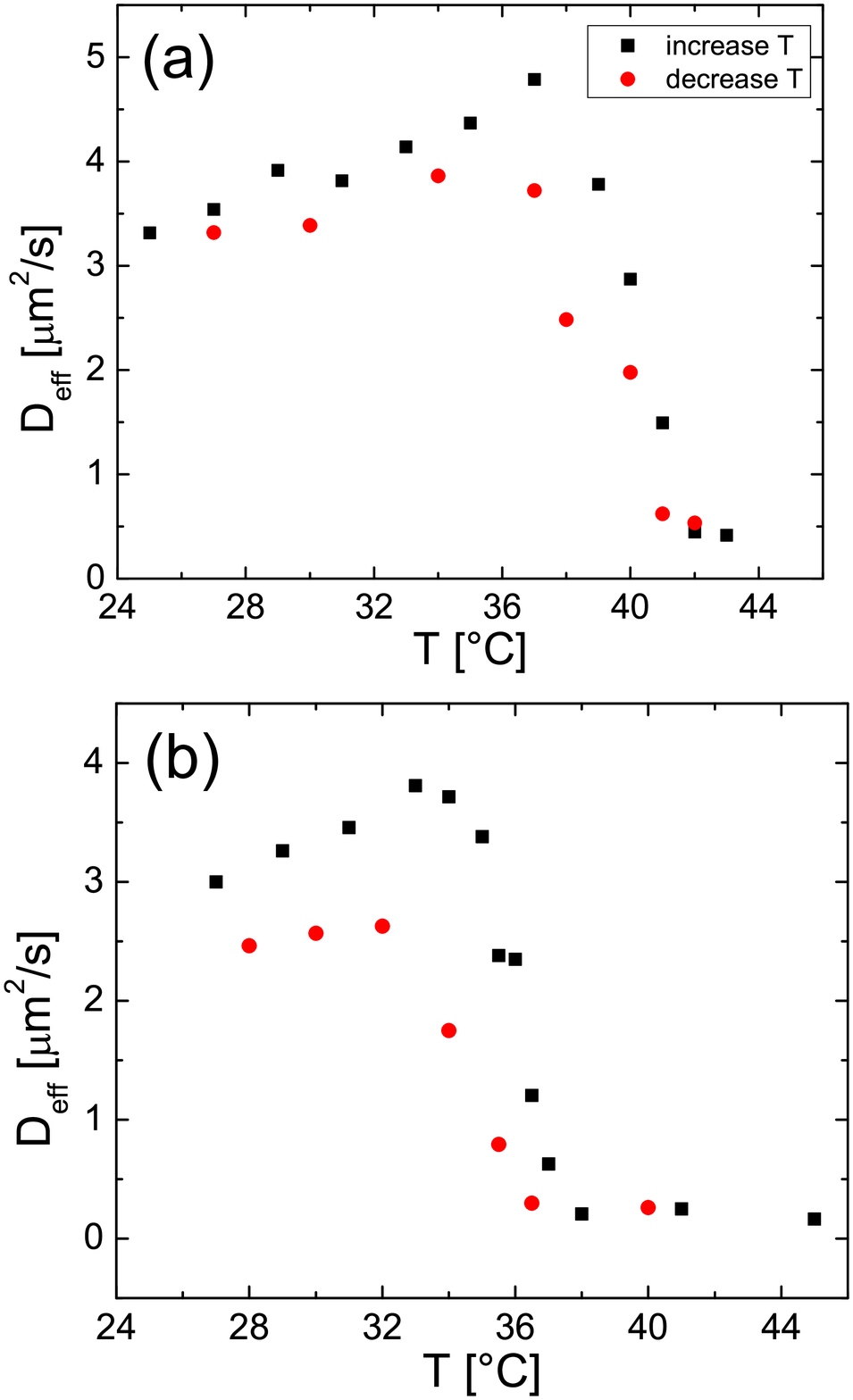,width=8.6cm} \caption{
    Diffusion coefficients of fd-PNIPAM at 0.l5 mg/ml as functions of temperature. Ionic strength $I=$ (a) 13~mM, (b) 155~mM. These
    values are determined from the first cumulant of $G_E(q,t)$ using Eq.~(\ref{eq:Deff}).
}
\label{DLS}
\end{figure}

\begin{figure}
\epsfig{file=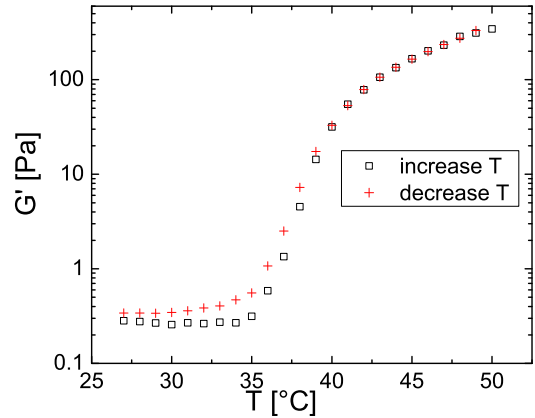,width=8.6cm} \caption{
    Reversibility of temperature-induced sol-gel transition. Measurements are made for fd-PNIPAM suspension at 8 mg/ml with increasing and decreasing temperature. The sample is oscillatorily probed at 1 Hz and the rate of temperature change is approximately $1^{\circ}$C/10 min in both directions.
}
\label{Treverse}
\end{figure}

\begin{figure}
\epsfig{file=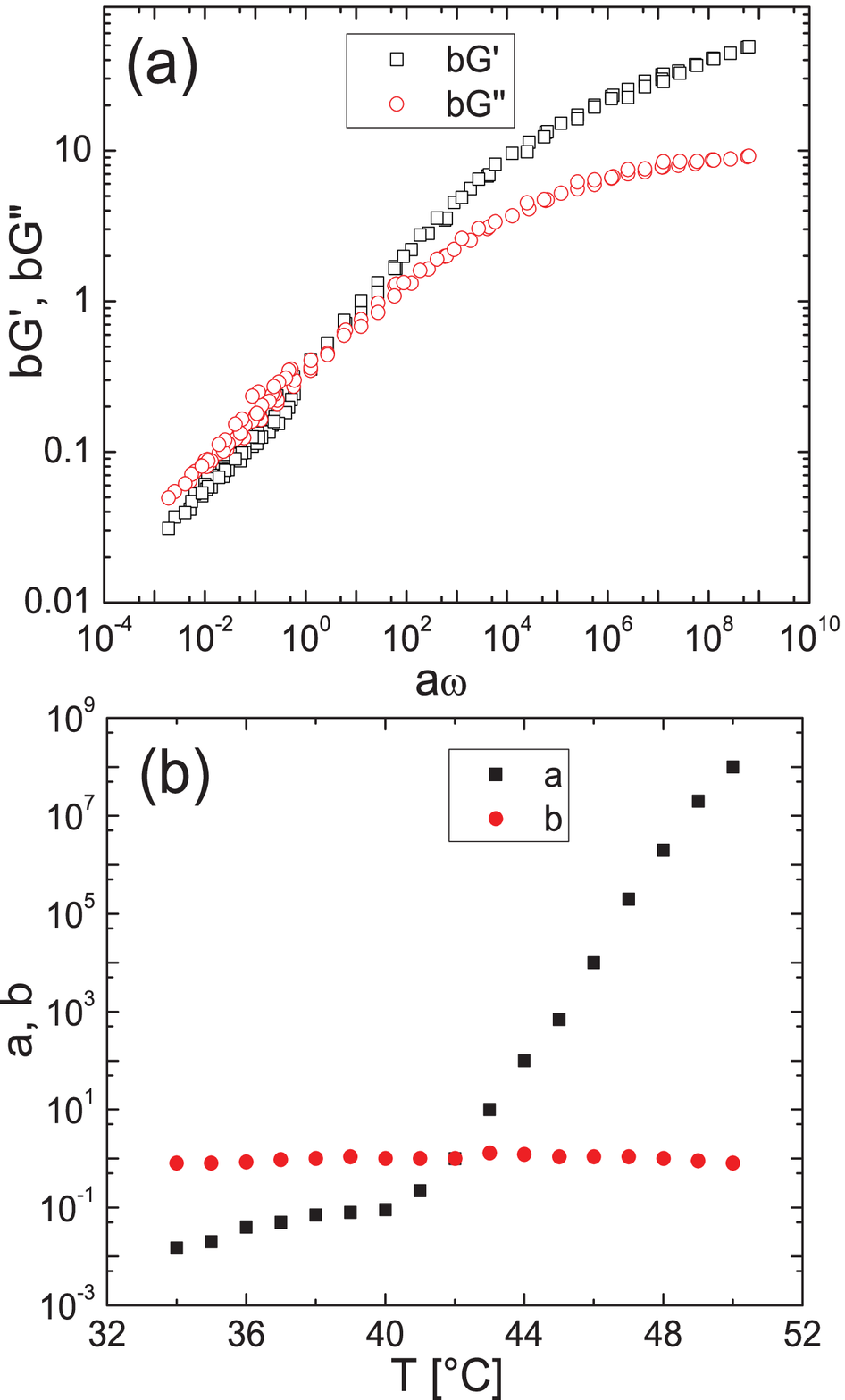,width=7.0cm} \caption{
    (a) Master curve showing scaled moduli as functions of scaled frequency. (b) Relationship between shift factors and temperature.  a: frequency shift factor. b: modulus shift factor.}
\label{master}
\end{figure}

%Rheometry
Fig.~\ref{fd&fd-PN} shows $G'(\omega)$ and $G''(\omega)$ for fd and fd-PNIPAM solutions measured at two different temperatures. Temperature change has little effect on the storage and loss moduli of fd suspension. By fitting the data to a power law, we have for fd $G'(\omega)\propto \omega^{0.9}$ and $G''(\omega)\propto \omega^{0.7}$. The frequency exponents are consistent with those measured with microrheology~\cite{Schmidt00}.
In contrast, fd-PNIPAM becomes solid-like at $38^{\circ}$C with $G'$ about five times $G''$. The linear moduli are nearly independent of frequency: $G'(\omega)\propto \omega^{0.14}$ and $G''(\omega)\propto \omega^{0.05}$.

%We investigate the effect of temperature on the rheological properties of the fd-PNIPAM network at two different ionic strengths as indicated in Fig.~\ref{fd-PNIPAM_coex}. As described earlier, electrostatic interaction dominates in the low salt regime (I = 10 mM), while steric repulsion dominates in the high salt regime (I = 150 mM). Before each measurement the suspension is allowed to equilibrate in the rheometer cell for five minutes.

We investigate the effect of ionic strength on the gelling process of the fd-PNIPAM network.
Fig.~\ref{fSweep} illustrates the frequency-dependent viscoelastic moduli as a function of temperature. Parts~(a) and~(b) represent data taken near the gel point $T=T_c$ from samples under low and high salt conditions respectively. For $T<T_c$. the suspension shows characteristics typical of a viscous fluid. The gel point is identified as the temperature at which $G'(\omega)$ and $G''(\omega)$ assume the same power law dependence on oscillation  frequency~\cite{Winter86}. As the temperature increases beyond $T=T_c$, both $G'(\omega)$ and $G''(\omega)$ increase dramatically and the suspension is clearly gel-like with $G'(\omega)$ weakly dependent on $\omega$.

The data in Fig.~\ref{fSweep} shows the high and low salt suspensions reach the gel point at different temperatures with the same power law slope. The sample at low ionic strength solidifies at $T_c=41^{\circ}$C, which is significantly greater than the $35^{\circ}$C gelling temperature for the sample at high ionic strength. However, both suspensions exhibit the same power law exponent $n=0.40\pm 0.02$ at the gel point.

As a check for the gel point, dynamic light scattering is performed on the fd-PNIPAM suspensions as shown in Fig.~\ref{DLS}. The onsets of aggregation for the low and high ionic strengths occur at 41 C and 36 C, respectively. Gelation occurs at the same temperatures as determined by light scattering and rheology. This ionic strength dependence of the gelation temperature arises from the fact that lowering solution ionic strength increases the electrostatic interaction between the rods, and therefore a larger attraction from the PNIPAM is required to induce aggregation.

To test the reversibility of the temperature-induced sol-gel transition, measurements are carried out on $G'$ at a frequency with increasing and decreasing temperature (Fig.~\ref{Treverse}). A slight hysteresis is found during the temperature sweep.

The storage and loss moduli curves at different temperatures can be scaled onto master curves. Through a procedure called time-temperature superposition (TTS)~\cite{Ferry80}, the $G'$ and $G''$ curves measured at different temperatures can be superposed by shifting along the logarithmic frequency and modulus axis. TTS enables one to probe viscoelasticity for a much larger frequency range than that experimentally accessible. The master curve as shown in Fig.~\ref{master}a reveals that fd-PNIPAM suspensions are a thermo-rheologically simple fluid, which means a variation in temperature corresponds to a shift in time scale~\cite{Ferry80}.
The rheological behaviors of fd-PNIPAM are reminiscent of those of polyethylene melts~\cite{Laun78}. At  high frequency, which corresponds to  high temperature $G'$ approaches a plateau value and is much larger than $G''$. In the low frequency limit, the suspension behaves like a fluid. $G'$ and $G''$ cross at an intermediate frequency with a slope of 0.36. The temperature-dependent shift factors are plotted in Fig.~\ref{master}b. Notably, the frequency shift factor exhibits a break of slope, signifying a phase transition, while there is only a minor shift along the logarithmic modulus axis.

Materials that are solid at high frequency and liquid like at low frequency are called thixotropic~\cite{Barnes97}. This is in stark contrast to colloidal gels of spherical particles whose fractal like microscopic structure leads to the opposite rheological behavior; fluid like at high frequency and solid like at low frequency~\cite{Trappe00}.

\section{Conclusion}
We have presented studies of a system of colloidal rods (fd) coated with the temperature-sensitive polymer (PNIPAM). At room temperature and high ionic strength, quantitative measurements of the I-N transition show fd-PNIPAM behaves as a sterically stabilized suspension. An increase in temperature, or equivalently, strength of attraction, does not lead to a widening of the coexistence concentration as expected. Instead a sol-gel transition arises, which we attribute to the collapse of the grafted PNIPAM polymers. Dynamic Light Scatterring  and rheometry demonstrate that the gelling process is reversible and ionic strength dependant. Furthermore, the rheological master curves for samples of different temperatures show that the fd-PNIPAM suspensions are rheologically similar to simple polymeric melts.

\section{Acknowledgement}
We thank D. A. Weitz, N. J. Wagner and Y. Hu for helpful discussions. Financial support of this work comes from NSF (DMR-0444172) and NSF-MRSEC (DMR-0820492).

%\bibliographystyle{apsrev}
%\bibliography{fd-PN_ref}

\begin{thebibliography}{27}
\expandafter\ifx\csname natexlab\endcsname\relax\def\natexlab#1{#1}\fi
\expandafter\ifx\csname bibnamefont\endcsname\relax
  \def\bibnamefont#1{#1}\fi
\expandafter\ifx\csname bibfnamefont\endcsname\relax
  \def\bibfnamefont#1{#1}\fi
\expandafter\ifx\csname citenamefont\endcsname\relax
  \def\citenamefont#1{#1}\fi
\expandafter\ifx\csname url\endcsname\relax
  \def\url#1{\texttt{#1}}\fi
\expandafter\ifx\csname urlprefix\endcsname\relax\def\urlprefix{URL }\fi
\providecommand{\bibinfo}[2]{#2}
\providecommand{\eprint}[2][]{\url{#2}}

\bibitem[{\citenamefont{Onsager}({1949})}]{Onsager49}
\bibinfo{author}{\bibfnamefont{L.}~\bibnamefont{Onsager}},
  \bibinfo{journal}{{Ann. N.Y. Acad. Sci.}} \textbf{\bibinfo{volume}{{51}}},
  \bibinfo{pages}{{627}} (\bibinfo{year}{{1949}}).

\bibitem[{\citenamefont{Davidson and Gabriel}({2005})}]{Davidson05}
\bibinfo{author}{\bibfnamefont{P.}~\bibnamefont{Davidson}} \bibnamefont{and}
  \bibinfo{author}{\bibfnamefont{J.~C.~P.} \bibnamefont{Gabriel}},
  \bibinfo{journal}{{Curr. Opin. Colloid Interface Sci.}}
  \textbf{\bibinfo{volume}{{9}}}, \bibinfo{pages}{{377}}
  (\bibinfo{year}{{2005}}).

\bibitem[{\citenamefont{Dogic and Fraden}({2006})}]{Dogic06}
\bibinfo{author}{\bibfnamefont{Z.}~\bibnamefont{Dogic}} \bibnamefont{and}
  \bibinfo{author}{\bibfnamefont{S.}~\bibnamefont{Fraden}},
  \bibinfo{journal}{{Curr. Opin. Colloid Interface Sci.}}
  \textbf{\bibinfo{volume}{{11}}}, \bibinfo{pages}{{47}}
  (\bibinfo{year}{{2006}}).

\bibitem[{\citenamefont{Asakura and Oosawa}({1958})}]{Asakura58}
\bibinfo{author}{\bibfnamefont{S.}~\bibnamefont{Asakura}} \bibnamefont{and}
  \bibinfo{author}{\bibfnamefont{F.}~\bibnamefont{Oosawa}},
  \bibinfo{journal}{{J. Polym. Sci.}} \textbf{\bibinfo{volume}{{33}}},
  \bibinfo{pages}{{183}} (\bibinfo{year}{{1958}}).

\bibitem[{\citenamefont{Warren}({1994})}]{Warren94}
\bibinfo{author}{\bibfnamefont{P.~B.} \bibnamefont{Warren}},
  \bibinfo{journal}{{J. Phys. I}} \textbf{\bibinfo{volume}{{4}}},
  \bibinfo{pages}{{237}} (\bibinfo{year}{{1994}}).

\bibitem[{\citenamefont{Lekkerkerker and Stroobants}({1994})}]{Lekkerkerker94}
\bibinfo{author}{\bibfnamefont{H.~N.~W.} \bibnamefont{Lekkerkerker}}
  \bibnamefont{and}
  \bibinfo{author}{\bibfnamefont{A.}~\bibnamefont{Stroobants}},
  \bibinfo{journal}{{Nuovo Cimento D}} \textbf{\bibinfo{volume}{{16}}},
  \bibinfo{pages}{{949}} (\bibinfo{year}{{1994}}).

\bibitem[{\citenamefont{Bolhuis et~al.}({1997})\citenamefont{Bolhuis,
  Stroobants, Frenkel, and Lekkerkerker}}]{Bolhuis97}
\bibinfo{author}{\bibfnamefont{P.~G.} \bibnamefont{Bolhuis}},
  \bibinfo{author}{\bibfnamefont{A.}~\bibnamefont{Stroobants}},
  \bibinfo{author}{\bibfnamefont{D.}~\bibnamefont{Frenkel}}, \bibnamefont{and}
  \bibinfo{author}{\bibfnamefont{H.~N.~W.} \bibnamefont{Lekkerkerker}},
  \bibinfo{journal}{{J. Chem. Phys.}} \textbf{\bibinfo{volume}{{107}}},
  \bibinfo{pages}{{1551}} (\bibinfo{year}{{1997}}).

\bibitem[{\citenamefont{Buitenhuis et~al.}({1995})\citenamefont{Buitenhuis,
  Donselaar, Buining, Stroobants, and Lekkerkerker}}]{Buitenhuis95}
\bibinfo{author}{\bibfnamefont{J.}~\bibnamefont{Buitenhuis}},
  \bibinfo{author}{\bibfnamefont{L.~N.} \bibnamefont{Donselaar}},
  \bibinfo{author}{\bibfnamefont{P.~A.} \bibnamefont{Buining}},
  \bibinfo{author}{\bibfnamefont{A.}~\bibnamefont{Stroobants}},
  \bibnamefont{and} \bibinfo{author}{\bibfnamefont{H.~N.~W.}
  \bibnamefont{Lekkerkerker}}, \bibinfo{journal}{{J. Colloid Interface Sci.}}
  \textbf{\bibinfo{volume}{{175}}}, \bibinfo{pages}{{46}}
  (\bibinfo{year}{{1995}}).

\bibitem[{\citenamefont{van Bruggen and Lekkerkerker}({2000})}]{Bruggen00}
\bibinfo{author}{\bibfnamefont{M.~P.~B.} \bibnamefont{van Bruggen}}
  \bibnamefont{and} \bibinfo{author}{\bibfnamefont{H.~N.~W.}
  \bibnamefont{Lekkerkerker}}, \bibinfo{journal}{{Macromolecules}}
  \textbf{\bibinfo{volume}{{33}}}, \bibinfo{pages}{{5532}}
  (\bibinfo{year}{{2000}}).

\bibitem[{\citenamefont{Dogic et~al.}({2004})\citenamefont{Dogic, Purdy,
  Grelet, Adams, and Fraden}}]{Dogic04}
\bibinfo{author}{\bibfnamefont{Z.}~\bibnamefont{Dogic}},
  \bibinfo{author}{\bibfnamefont{K.}~\bibnamefont{Purdy}},
  \bibinfo{author}{\bibfnamefont{E.}~\bibnamefont{Grelet}},
  \bibinfo{author}{\bibfnamefont{M.}~\bibnamefont{Adams}}, \bibnamefont{and}
  \bibinfo{author}{\bibfnamefont{S.}~\bibnamefont{Fraden}},
  \bibinfo{journal}{{Phys. Rev. E}} \textbf{\bibinfo{volume}{{69}}}
  (\bibinfo{year}{{2004}}).

\bibitem[{\citenamefont{Flory}({1956})}]{Flory56}
\bibinfo{author}{\bibfnamefont{P.~J.} \bibnamefont{Flory}},
  \bibinfo{journal}{{Proc. R. Soc. London Ser. A}}
  \textbf{\bibinfo{volume}{{234}}}, \bibinfo{pages}{{73}}
  (\bibinfo{year}{{1956}}).

\bibitem[{\citenamefont{Miller et~al.}({1974})\citenamefont{Miller, Wu, Wee,
  Santee, Rai, and Goebel}}]{Miller74}
\bibinfo{author}{\bibfnamefont{W.~G.} \bibnamefont{Miller}},
  \bibinfo{author}{\bibfnamefont{C.~C.} \bibnamefont{Wu}},
  \bibinfo{author}{\bibfnamefont{E.~L.} \bibnamefont{Wee}},
  \bibinfo{author}{\bibfnamefont{G.~L.} \bibnamefont{Santee}},
  \bibinfo{author}{\bibfnamefont{J.~H.} \bibnamefont{Rai}}, \bibnamefont{and}
  \bibinfo{author}{\bibfnamefont{K.~G.} \bibnamefont{Goebel}},
  \bibinfo{journal}{{Pure Appl. Chem.}} \textbf{\bibinfo{volume}{{38}}},
  \bibinfo{pages}{{37}} (\bibinfo{year}{{1974}}).

\bibitem[{\citenamefont{Tang and Fraden}(1995)}]{Tang95}
\bibinfo{author}{\bibfnamefont{J.~X.} \bibnamefont{Tang}} \bibnamefont{and}
  \bibinfo{author}{\bibfnamefont{S.}~\bibnamefont{Fraden}},
  \bibinfo{journal}{Liq. Cryst.} \textbf{\bibinfo{volume}{19}},
  \bibinfo{pages}{459} (\bibinfo{year}{1995}).

\bibitem[{\citenamefont{de~Gennes}(1979)}]{Gennes79}
\bibinfo{author}{\bibfnamefont{P.~G.} \bibnamefont{de~Gennes}},
  \emph{\bibinfo{title}{Scaling Concepts In Polymer Physics}}
  (\bibinfo{publisher}{Cornell University Press, Ithaca, New York},
  \bibinfo{year}{1979}).

\bibitem[{\citenamefont{Schild}({1992})}]{Schild92}
\bibinfo{author}{\bibfnamefont{H.~G.} \bibnamefont{Schild}},
  \bibinfo{journal}{{Prog. Polym. Sci.}} \textbf{\bibinfo{volume}{{17}}},
  \bibinfo{pages}{{163}} (\bibinfo{year}{{1992}}).

\bibitem[{\citenamefont{Fraden}(1995)}]{Fraden95}
\bibinfo{author}{\bibfnamefont{S.}~\bibnamefont{Fraden}},
  \emph{\bibinfo{title}{$\mathrm{in}$ Observation, Prediction, and Simulation
  of Phase Transitions in Complex Fluids}}, edited by M. Baus, L.F. Rull, and
  J.P. Ryckaert (\bibinfo{publisher}{Kluwer Academic, Dordrecht},
  \bibinfo{year}{1995}), \bibinfo{note}{pp. 113-164}.

\bibitem[{\citenamefont{Khalil et~al.}({2007})\citenamefont{Khalil, Ferrer,
  Brau, Kottmann, Noren, Lang, and Belcher}}]{Khalil07}
\bibinfo{author}{\bibfnamefont{A.~S.} \bibnamefont{Khalil}},
  \bibinfo{author}{\bibfnamefont{J.~M.} \bibnamefont{Ferrer}},
  \bibinfo{author}{\bibfnamefont{R.~R.} \bibnamefont{Brau}},
  \bibinfo{author}{\bibfnamefont{S.~T.} \bibnamefont{Kottmann}},
  \bibinfo{author}{\bibfnamefont{C.~J.} \bibnamefont{Noren}},
  \bibinfo{author}{\bibfnamefont{M.~J.} \bibnamefont{Lang}}, \bibnamefont{and}
  \bibinfo{author}{\bibfnamefont{A.~M.} \bibnamefont{Belcher}},
  \bibinfo{journal}{{Proc. Natl. Acad. Sci.}} \textbf{\bibinfo{volume}{{104}}},
  \bibinfo{pages}{{4892}} (\bibinfo{year}{{2007}}).

\bibitem[{\citenamefont{Maniatis et~al.}(1989)\citenamefont{Maniatis, Sambrook,
  and Fritsch}}]{Maniatis89}
\bibinfo{author}{\bibfnamefont{T.}~\bibnamefont{Maniatis}},
  \bibinfo{author}{\bibfnamefont{J.}~\bibnamefont{Sambrook}}, \bibnamefont{and}
  \bibinfo{author}{\bibfnamefont{E.~F.} \bibnamefont{Fritsch}},
  \emph{\bibinfo{title}{Molecular Cloning: A Laboratory Manual}}
  (\bibinfo{publisher}{Cold Spring Harbor Laboratory Press, Plainview, NY},
  \bibinfo{year}{1989}), \bibinfo{edition}{2nd} ed.

\bibitem[{\citenamefont{Grelet and Fraden}({2003})}]{Grelet03}
\bibinfo{author}{\bibfnamefont{E.}~\bibnamefont{Grelet}} \bibnamefont{and}
  \bibinfo{author}{\bibfnamefont{S.}~\bibnamefont{Fraden}},
  \bibinfo{journal}{Phys. Rev. Lett.} \textbf{\bibinfo{volume}{{90}}},
  \bibinfo{pages}{198302} (\bibinfo{year}{{2003}}).

\bibitem[{\citenamefont{Berne and Pecora}(1976)}]{Berne76}
\bibinfo{author}{\bibfnamefont{J.~B.} \bibnamefont{Berne}} \bibnamefont{and}
  \bibinfo{author}{\bibfnamefont{R.}~\bibnamefont{Pecora}},
  \emph{\bibinfo{title}{Dynamic Light Scattering}} (\bibinfo{publisher}{Wiley,
  New York}, \bibinfo{year}{1976}).

\bibitem[{\citenamefont{Dogic and Fraden}({2001})}]{Dogic01}
\bibinfo{author}{\bibfnamefont{Z.}~\bibnamefont{Dogic}} \bibnamefont{and}
  \bibinfo{author}{\bibfnamefont{S.}~\bibnamefont{Fraden}},
  \bibinfo{journal}{{Philos. Trans. R. Soc. London, Ser. A}}
  \textbf{\bibinfo{volume}{{359}}}, \bibinfo{pages}{{997}}
  (\bibinfo{year}{{2001}}).

\bibitem[{\citenamefont{Schmidt et~al.}({2000})\citenamefont{Schmidt, Hinner,
  Sackmann, and Tang}}]{Schmidt00}
\bibinfo{author}{\bibfnamefont{F.~G.} \bibnamefont{Schmidt}},
  \bibinfo{author}{\bibfnamefont{B.}~\bibnamefont{Hinner}},
  \bibinfo{author}{\bibfnamefont{E.}~\bibnamefont{Sackmann}}, \bibnamefont{and}
  \bibinfo{author}{\bibfnamefont{J.~X.} \bibnamefont{Tang}},
  \bibinfo{journal}{{Phys. Rev. E}} \textbf{\bibinfo{volume}{{62}}},
  \bibinfo{pages}{{5509}} (\bibinfo{year}{{2000}}).

\bibitem[{\citenamefont{Winter and Chambon}({1986})}]{Winter86}
\bibinfo{author}{\bibfnamefont{H.~H.} \bibnamefont{Winter}} \bibnamefont{and}
  \bibinfo{author}{\bibfnamefont{F.}~\bibnamefont{Chambon}},
  \bibinfo{journal}{{J. Rheol.}} \textbf{\bibinfo{volume}{{30}}},
  \bibinfo{pages}{{367}} (\bibinfo{year}{{1986}}).

\bibitem[{\citenamefont{Ferry}(1980)}]{Ferry80}
\bibinfo{author}{\bibfnamefont{J.~D.} \bibnamefont{Ferry}},
  \emph{\bibinfo{title}{Viscoelastic Properties of Polymers}}
  (\bibinfo{publisher}{Wiley, New York}, \bibinfo{year}{1980}),
  \bibinfo{edition}{3rd} ed.

\bibitem[{\citenamefont{Laun}({1978})}]{Laun78}
\bibinfo{author}{\bibfnamefont{H.~M.} \bibnamefont{Laun}},
  \bibinfo{journal}{{Rheol. Acta}} \textbf{\bibinfo{volume}{{17}}},
  \bibinfo{pages}{{1}} (\bibinfo{year}{{1978}}).

\bibitem[{\citenamefont{Barnes}({1997})}]{Barnes97}
\bibinfo{author}{\bibfnamefont{H.~A.} \bibnamefont{Barnes}},
  \bibinfo{journal}{{J. Non-Newtonian Fluid Mech.}}
  \textbf{\bibinfo{volume}{{70}}}, \bibinfo{pages}{{1}}
  (\bibinfo{year}{{1997}}).

\bibitem[{\citenamefont{Trappe and Weitz}({2000})}]{Trappe00}
\bibinfo{author}{\bibfnamefont{V.}~\bibnamefont{Trappe}} \bibnamefont{and}
  \bibinfo{author}{\bibfnamefont{D.~A.} \bibnamefont{Weitz}},
  \bibinfo{journal}{Phys. Rev. Lett.} \textbf{\bibinfo{volume}{{85}}},
  \bibinfo{pages}{{449}} (\bibinfo{year}{{2000}}).

\end{thebibliography}

\end{document}